\documentclass[12pt]{article}
\usepackage{amssymb,amsmath}
\usepackage[noblocks]{authblk}
\usepackage[top=0.75in, bottom=0.75in, left=0.75in, right=0.75in, dvips]{geometry}
\usepackage{caption}
\begin{document}
\textwidth 10.0in
\textheight 9.0in
\topmargin -0.60in
\title{Treatment of a System with Explicitly Broken Gauge Symmetries}
\author[3]{F.A. Chishtie} \author[2]{T. Hanif} \author[1,2]{D.G.C. McKeon}
\affil[1] {Department of Applied Mathematics, The
University of Western Ontario, London, ON N6A 5B7, Canada}
\affil[2] {Department of Mathematics and
Computer Science, Algoma University, Sault St.Marie, ON P6A
2G4, Canada}
\affil[3] {Department of Theoretical Physics, University of Dhaka, Dhaka-1000, Bangladesh}
\affil[4] {Department of Space Science, Institute of Space Technology, Islamabad 44000, Pakistan}
\maketitle

\begin{abstract}
A system in which the free part of the action possesses a gauge symmetry that is not respected by the interacting part presents problems when quantized. We illustrate how the Dirac constraint formalism can be used to address this difficulty by considering an antisymmetric tensor field interacting with a spinor field.
\end{abstract}

\section{Introduction}

The Fadeev-Popov (FP) approach to the quantization of massless Yang-Mills (YM) gauge field [1] is quite useful. It provides a way of eliminating non-physical degrees of freedom that are merely gauge artifacts while allowing for the introduction of a variety of gauge choices, both covariant and non-covariant. (In fact, it is possible extend the FP procedure to accommodate more than one gauge-fixing condition [2,3].)

A practical problem overcome by the FP technique is the difficulty in obtaining the free-field propagator for a gauge field. Naively, the propagator for a massless vector gauge field $V_{\mu}$ involves inverting the operator $(\partial^2 g_{\mu\nu}-\partial_{\mu}\partial_{\nu})$, but this is impossible because the gauge invariance $V_{\mu}\rightarrow V_{\mu}+ \partial_{\mu}{\theta}$ means that this operator has a vanishing eigenvalue. In the FP approach, such bilinears are supplemented by a gauge breaking term such as $-\frac{1}{2}\left({\partial\cdot V}\right)^2$ making it possible to obtain the propagator. Of course, if the classical action also has a term that explicitly breaks gauge invariance (such as $\frac{1}{2} m^2 V_{\mu}V^{\mu}$) this problem does not arise.

However, it is possible to have a gauge invariance present in the bilinear part of the Lagrangian that is broken explicitly by the interaction. (For example, the interaction $-\lambda(V_{\mu}V^{\mu})^2$ could occur in addition to the Maxwell action for $V_{\mu}$.) In this case, the FP procedure is not directly applicable, and yet the free field propagator cannot be obtained from the bilinear part of the action as by itself it possesses a gauge invariance.

In order to address this problem, it is necessary to keep in mind that the FP procedure is equivalent to the path integral (PI) as derived from canonical quantization for YM gauge theories [4,5] but that is not always the case [6,7]. A system that involves first and/or second class constraints (as introduced by Dirac [8,9]) and thereby possesses a gauge invariance that can only be quantized using the PI if the measure of the PI is modified by the appropriate functional determinants and delta functions [4,10]. These modifications are equivalent to having the FP measure for the PI for YM theory, but this need not always be the case.

Recalling this, we examine the problem of quantizing an antisymmetric tensor field $\phi_{\mu\nu}$ interacting with a spinor $\psi$ through a magnetic moment interaction. We consider both the massless, gauge invariant, free field action for $\phi_{\mu\nu}$, and also supplement it with a scalar and/or pseudoscalar mass term. If these mass terms vanish, we encounter the problem mentioned above of defining the free propagator when the interaction is not gauge invariant. It is shown that this model has constraints that modify the measure of the PI so that the functional integral is well defined and there is a free field propagator for $\phi_{\mu\nu}$.

An unresolved problem remains however; it is not clear if the resulting PI is covariant as manifest covariance has been lost. The difficulty originally plagued both quantum electrodynamics [11-14] and YM theory [15, 16] but in these theories the FP approach made it possible to retain manifest covariance.

We use the notation outlined in the appendix.

\section{A Spinor-Tensor Model}
The action
\[ \mathcal{L}_{\phi} = \frac{1}{12} \left( \partial_{\mu}\phi_{\nu\lambda}+\partial_{\nu}\phi_{\lambda\mu}+\partial_{\lambda}\phi_{\mu\nu} \right)^2\equiv G_{\lambda\mu\nu}^2\eqno(1) \]
for the field $\phi_{\mu\nu}=-\phi_{\nu\mu}$, possesses the gauge invariance
\[ \delta\phi_{\mu\nu}=\partial_{\mu}\theta_{\nu}-\partial_{\nu}\theta_{\mu}\eqno(2) \]
Consequently, if we write
\[ \mathcal{L}_{\phi} = \frac{1}{2}\phi_{\alpha\beta} \left(-\frac{1}{2} \partial^2I^{\alpha\beta,\gamma\delta}+Q^{\alpha\beta,\gamma\delta} \right)\phi_{\gamma\delta}\eqno(3) \]
where
\[ I^{\alpha\beta,\gamma\delta}=\frac{1}{2}\left(g^{\alpha\gamma}g^{\beta\delta}-g^{\alpha\delta}g^{\beta\gamma}\right)\eqno(4a) \]
\[ Q^{\alpha\beta,\gamma\delta}=\frac{1}{4}\left(\partial^{\alpha\gamma}g^{\beta\delta}-\partial^{\beta\gamma}g^{\alpha\delta}
+\partial^{\beta\delta}g^{\alpha\gamma}-\partial^{\alpha\delta}g^{\beta\gamma}\right)\eqno(4b) \]
we find that
\[ M_0^{\alpha\beta,\gamma\delta}=-\frac{1}{2}\partial^2I^{\alpha\beta,\gamma\delta}+Q^{\alpha\beta,\gamma\delta}\eqno(5) \] $M_0^{\alpha\beta,\gamma\delta}\partial_{\gamma}=0$ and thus $M_0^{\alpha\beta,\gamma\delta}$ has no inverse. We can supplement $M_0$ with
\[ M_{\mu^2}^{\alpha\beta,\gamma\delta}=-\frac{\mu^2}{4}\epsilon^{\alpha\beta\gamma\delta}\eqno(6a) \]
and/or
\[ M_{m^2}^{\alpha\beta,\gamma\delta}=-\frac{m^2}{2}I^{\alpha\beta,\gamma\delta} \eqno(6b) \]
and it is obvious that since neither of these are invariant under that transformation of eq.(2), one can now find a free propagator for $\phi_{\mu\nu}$.

However, if we simply take $\mathcal{L}_{\phi}$ and couple $\phi_{\mu\nu}$ to a spinor $\psi$ so that $\mathcal{L}=\mathcal{L}_{\phi}+\mathcal{L}_{\psi}$ where
\[ \mathcal{L}_{\psi} = \overline{\psi} \left(i\gamma\cdot\partial+g\sigma^{\mu\nu}\gamma^5\phi_{\mu\nu}\right)\psi\eqno(7) \]
then the free Lagrangian for $\phi_{\mu\nu}$ is gauge invariant while the interaction with $\psi$ is not and the problem outlined in the preceding section occurs.

We now recall that if one employs canonical quantization for a system with first class constraints $\varphi_i$, second class constraints $\theta_i$ and gauge conditions $\gamma_i$, then the transition amplitude is given by the PI
\[ <out|in> = \int dq_i dp_i M \exp i\int_{-\infty}^{\infty}dt\left(\dot{q_i}p_i-H(q_i,p_i)\right)\eqno(8) \]
where $H$ is the canonical Hamiltonian, $q_i(t\rightarrow\pm\infty)=(q_{out},q_{in})$ and $M$ is the contribution to the functional measure that is a consequence of constraints being present [4,10]
\[ M = \delta(\phi_i)\delta(\theta_i)\delta(\gamma_i) det\left\lbrace\phi_i,\gamma_j\right\rbrace det^{1/2}\left\lbrace\theta_i,\theta_j\right\rbrace\eqno(9) \]
with $\lbrace,\rbrace$ denoting the Poisson Bracket (PB).

For YM theory, there is a single gauge invariance and has been shown [4,5] that for this case the measure of eq.(9) is the same as the FP measure. However, in other cases (such as the non-Abelian extension of $\mathcal{L}_{\phi}$ [6], the first order Einstein-Hilbert action in $d\geq3$ dimensions [7] and supergravity in $2+1$ dimensions [17]) this equivalence does not hold.

We are thus motivated to study the constraint structure of $\mathcal{L}_{\phi}+\mathcal{L}_{\psi}$ (possibly supplemented by
\[\mathcal{L}_{\mu^2}=-\frac{\mu^2}{8}\epsilon^{\mu\nu\lambda\sigma}\phi_{\mu\nu}\phi_{\lambda\sigma}\eqno(10a) \]
and/or
\[ \mathcal{L}_{m^2}=-\frac{m^2}{4}\phi^{\mu\nu}\phi_{\mu\nu}). \eqno(10b) \]
in order to see how a suitable transition amplitude can be defined by using a PI.

We begin by defining
\[ A_i=\phi_{0i},\quad B_i = \frac{1}{2}\epsilon_{ijk}\phi_{jk} \eqno(11a,b) \]
so that
\[ \mathcal{L} =\frac{1}{2}\dot{B}_i\dot{B}_i-\epsilon_{ijk}A_i\partial_j\dot{B}_k
+\frac{1}{2}A_i\left(\partial_i\partial_j-\partial^2\delta_{ij}\right)A_j - \frac{1}{2}(B_{i,i})^2 \eqno(12) \]
\[ \hspace{1cm}- \mu^2 A_iB_i + \frac{m^2}{2}(A_i^2-B_i^2)+i\psi^{\dag}(\dot{\psi}+\alpha^i\psi_{,i})+g\psi^{\dag}(S^iA_i+i\gamma^iB_i)\psi\nonumber \]
From eq. (12) it is apparent that the canonical momentum associated with the fields $A_i$, $B_i$, $\psi$ and $\psi^{\dag}$ are respectively
\[ \pi_i^A=0 \eqno(13a) \]
\[ \pi_i^B=\dot{B}_i-\epsilon_{ijk}\partial_j A_k \eqno(13b) \]
\[ \pi^{\dag}=-i\psi^{\dag} \eqno(13c) \]
\[ \pi=0 \eqno(13d) \]
Eqs. (13 a,c,d) are primary constraints. Since by eq. (A6.a)
\[ \left\lbrace \pi^{\dag}+i\psi^{\dag},\pi\right\rbrace=-i \eqno(14) \]
we see that there are two primary second class constraints
\[ \chi_1=\pi^{\dag}+i\psi^{\dag}\eqno(15a) \]
\[ \chi_2=\pi. \eqno(15b) \]
The canonical Hamiltonian is now given by

\[ \mathcal{L} =\frac{1}{2}\pi_i^{B}\pi_i^{B}+\epsilon_{ijk}\pi_i^{B}\partial_jA_k
+ \frac{1}{2}(B_{i,i})^2+\mu^2 A_iB_i-\frac{m^2}{2}(A_i^2-B_i^2) \eqno(16) \]
\[ \hspace{1cm} -i\psi^{\dag}\alpha^i\psi_{,i}-g\psi^{\dag}(S^iA_i+i\gamma^iB_i)\psi.\nonumber \]
In order to eliminate the two second class constraints of eq.(15) we define the Dirac Bracket (DB)
\[ \left\lbrace X,Y\right\rbrace^{*}=\left\lbrace X,Y\right\rbrace-i\left[\left\lbrace X,\chi_1\right\rbrace\left\lbrace \chi_2,Y\right\rbrace+\left\lbrace X,\chi_2\right\rbrace\left\lbrace \chi_1,Y\right\rbrace\right] \eqno(17) \]
if $X$ and $Y$ are Fermionic, so that
\[ \left\lbrace \psi,\psi^{\dag}\right\rbrace^{*}=-i \eqno(18) \]

The primary constraint of eq. (13a) now leads to the secondary constraints
\[ \Lambda_i=\epsilon_{ijk}\partial_j\pi_k^B+\mu^2B_i-m^2A_i-g\psi^{\dag}S^i\psi. \eqno(19) \]

Consider first the limit $\mu^2=m^2=g=0$. In this case there are three secondary constraints
\[ \lambda_i=\epsilon_{ijk}\partial_j\pi_k^B \eqno(20) \]
but not all of them are independent as $\partial_i\lambda_i=0$. It is easily shown that there are no tertiary (third generation) constraints and that the constraints of eq.(13a) and (20) are all first class. With five first class constraints and five associated gauge conditions, there are ten constraints on the 12 variables in phase space ($\phi_{\mu\nu}$ and the associated momenta); we are left with $12-10=2$ physical degrees of freedom in phase space. The gauge generator of Henneaux, Teitelboim and Zanelli [18] is of the form
\[ G=\nu^i\pi_i^A+\mu^i\lambda_i \eqno(21) \]
and the equation
\[ \dot{\nu}^{i}\pi_i^A+\dot{\mu}^{i}\lambda_i+\left\lbrace G,\int dx\left(\mathcal{H}_c+U_i\pi_i^A\right)\right\rbrace -\delta U^i\pi_i^A=0 \eqno(22) \]
results in
\[ \nu^i=\dot{\mu}^{i}. \eqno(23) \]

We then find that the gauge generator $G$ will generate the transformation of eq.(2) with $\theta_i=\mu^i$ and $\theta_0=0$.

If we take $\mu^2=g=0$ in eq. (19), we have the secondary constraint
\[ \Lambda_i^{(m^2)}=\lambda_i-m^2A_i \eqno(24) \]
then eq.(13a) and (24) define a set of six second class constraints as
\[ \left\lbrace\pi_i^A,\Lambda_j^{(m^2)}\right\rbrace=m^2\delta_{ij} \eqno(25) \]

There are now six second class constraints on $\phi_{\mu\nu}$ and its conjugate momenta leaving $12-6=6$ physical degrees of freedom in phase space.

With $m^2=g=0$ in eq.(19), we then have the secondary constraint [6]
\[ \Lambda_{i}^{(\mu^2)}=\lambda_i+\mu^2B_i \eqno(26) \]
As $\left\lbrace\pi^A_i,\Lambda^{(\mu^2)}_j\right\rbrace=0$, it is necessary to check if there are any tertiary constraints; one easily finds that there is now the tertiary constraint
\[ T_{i}=\mu^2\pi_i^B \eqno(27) \]
We see that $\Lambda^{(\mu^2)}_i$ and $T_i$ are second class as
\[ \left\lbrace\Lambda^{(\mu^2)}_i,T_j\right\rbrace=\mu^4\delta_{ij} \eqno(28) \]
with six second class constraints, plus the three first class constraints $\pi^A_i$ and the associated gauge conditions, there are $6+3+3$ constraints in phase space on $\phi_{\mu\nu}$ and its canonical momenta.

This leaves no net degrees of freedom for the field $\phi_{\mu\nu}$, which is consistent with the results of refs. [19,20]. It is peculiar that adding a mass term reduces the number of degrees of freedom; this is unlike the addition of a Proca mass to vector gauge field $V_{\mu}$ in which case $V_{\mu}$ acquires a longitudinal polarization.

However, if we only take $m^2=0$ in eq.(19), then we have the constraint,
\[ L_i=\Lambda^{(\mu^2)}_i-g\psi^{\dag}S^i\psi. \eqno(29) \]
From eq.(18) it follows that
\[ \left\lbrace L_i,L_j\right\rbrace^*=g^2\psi^{\dag}S^i\left\lbrace\psi,\psi^{\dag}\right\rbrace^*S^j\psi+
 (S^i\psi)^{\dag}\left\lbrace\psi^{\dag},\psi\right\rbrace^*(\psi^{\dag}S^j)^{\dag}=
 2g^2\epsilon^{ijk}\left(\psi^{\dag}\Sigma^k\psi\right) \eqno(30) \]
However, this does not mean that all of the constraints $L_i$ are second class; the number of Bosonic second class constraints must be even and so we chose to designate $L_1$ and $L_2$ to be those second class constraints. (A similar situation arises when analyzing the canonical structure of the action for the superparticle in $2+1$ and $3+1$ dimensions [21].) With $L_1$ and $L_2$ being second class, we define a ``second stage DB" [22]
\[ \left\lbrace X,Y\right\rbrace^{**}=\left\lbrace X,Y\right\rbrace^{*}+\frac{1}{2\Delta}\left[\left\lbrace X,L_2\right\rbrace^{*}\left\lbrace L_1,Y\right\rbrace^{*}-\left\lbrace X,L_1\right\rbrace^{*}\left\lbrace L_2,Y\right\rbrace^{*}\right] \eqno(31) \]
where $\Delta=\psi^{\dag}\Sigma^3\psi$. We now consider the consequence for $L$ of having eliminated $L_1$ and $L_2$ through definition of this DB $\lbrace,\rbrace^{**}$. We find that
\[ \left\lbrace \Lambda_3,\int dx\mathcal{H}\right\rbrace^{**}=\frac{g^2}{\Delta}\left[\epsilon_{ijk}\left(\psi^{\dag}\Sigma^i\psi\right)\partial_j
\left(\psi^{\dag}\gamma^k\psi\right)+\psi^{\dag}S^i\psi\left(\psi^{\dag}S^i\alpha^j\psi_{,j}+\psi_{,j}^{\dag}\alpha^jS^i\psi
-2B_i\psi^{\dag}\gamma^5\psi\right)\right] \eqno(32) \]
\[\equiv\frac{g^2}{\Delta}K\]

We are thus confronted with a tertiary constraint $K$ if $g^2\neq0$; together $L_3$ and $K$ are obviously second class even if $\mu^2\neq0$. Consequently there are 28 degrees of freedom in phase space ($\phi_{\mu\nu}$, $\psi$, $\psi^{\dag}$ and their conjugate momenta), eight primary second class constraints ($\chi_1$ and $\chi_2$), three secondary second class constraints ($L_i$) and one tertiary second class constraint ($K$). If $m^2=0$, $A_i$ in eq.(16) acts as a Lagrange multiplier for the constraint $L_i$ and the arbitrariness is reflected in the fact that $\pi^A_i=0$ (eq. (13a)); in this case $\pi^A_i$ is first class constraint. With $8+3+1=12$ second class constraints, three first class constraints and three gauge conditions, there are $12+3+3=18$ restrictions on these 28 degrees of freedom. The ten physical degrees of freedom are the two polarizations of the spinor and anti-spinor, one degree associated with the tensor, plus the associated canonical momenta. Surprisingly, if $m^2=0$, $\mu^2\neq0$ a degree of freedom absent from $\phi_{\mu\nu}$ when $g=0$ is restored when $g\neq0$.

This can be seen by examining the equations of motion that follow from
\[ \mathcal{L} = \frac{1}{12} G_{\mu\nu\lambda}^2-\frac{\mu^2}{8}\epsilon^{\mu\nu\lambda\sigma}\phi_{\mu\nu}\phi_{\lambda\sigma}+\psi^{\dag}
\left(i\gamma\cdot\partial+g\sigma^{\mu\nu}\gamma^5\phi_{\mu\nu}\right)\psi\eqno(33) \]
The equation of motion for $\phi_{\mu\nu}$ that follows from eq. (33) is (with $J^{\mu\nu}=\psi^{\dag}\sigma^{\mu\nu}\gamma^5\psi$)
\[ -\frac{1}{2}\partial_{\mu}G^{\mu\nu\lambda}-\frac{\mu^2}{4}\epsilon^{\alpha\beta\nu\lambda}\phi_{\alpha\beta}
+gJ^{\nu\lambda}=0\eqno(34) \]
Upon operating on eq. (34) with $\partial_{\nu}$ we obtain
\[ \frac{\mu^2}{12}\epsilon^{\lambda\alpha\beta\gamma}G_{\alpha\beta\gamma}+g\partial_{\nu}J^{\nu\lambda}=0,\eqno(35) \]
which in turn implies that
\[ G_{\alpha\beta\gamma}=-\frac{2g}{\mu^2}\epsilon_{\lambda\alpha\beta\gamma}\partial_{\nu}J^{\nu\lambda}.\eqno(36) \]
If $g=0$, then by eqs.(34,36) $\phi_{\mu\nu}=0$; for $g\neq0$ these equations imply that
\[ \phi_{\mu\nu}=-\frac{2g}{\mu^2}\left[\frac{1}{\mu^2}\left(\partial_{\mu\rho}^2J^{\rho}_{\nu}-
\partial_{\nu\rho}^2J^{\rho}_{\mu}\right)-\epsilon_{\mu\nu\lambda\sigma}J^{\lambda\sigma}\right].\eqno(37) \]
showing that if $m^2=0$, $\mu^2\neq0$ then the tensor field is fixed by the spinor field.

Having examined the constrained structure of our tensor-spinor model when the possibility of having terms which break the symmetry of eq.(2) appear in the action, we are in a position to determine the contribution of $M$ in eq.(9) to the measure of the PI of eq.(8). We have seen that if $m^2=0$, then the constraint of eq.(13a) remains first class. Choosing the associated gauge fixing condition to be $A_i=0$ makes it possible to obtain a free field propagator for $\phi_{\mu\nu}$. If $m^2\neq0$ then the propagator is obtained from
\[ \left[-\frac{1}{2}(\partial^2+m^2)I^{\alpha\beta,\gamma\delta}+Q^{\alpha\beta,\gamma\delta}\right]^{-1}=
-\frac{2}{\partial^2+m^2}\left[I^{\alpha\beta,\gamma\delta}+\frac{2}{m^2}Q^{\alpha\beta,\gamma\delta}\right] \eqno(38) \]
which is well defined.

The contribution of second class constraints to $M$ in eq.(9) given by $det^{1/2}\lbrace\theta_i,\theta_j\rbrace$ is particularly complicated for $g\neq0$ in light of the form of $L_i$ and $K$ in eqs.(29,32).

It is not apparent that the final resulting expression for the transition amplitude in eq.(8) is covariant in view of the
fact that manifest covariance has clearly been lost. A similar problem occurs when applying eqs.(8,9) to the first-order Einstein-Hilbert action in $d\geq3$ dimensions [7] and the non-Abelian tensor field $\phi^a_{\mu\nu}$ when it has a pseudoscalar mass term [6]. In both cases, the contribution of the second class constraints to the measure of the path integral is non-trivial and lacks manifest covariance.

We have not considered the problem of renormalization for this system.

\section{Discussion}
We have shown that a PI can be well defined for a tensor field interacting with a spinor field provided full use is made of the Dirac constraints occurring in this system. However, as the second class constraints occurring are non-trivial, the PI is no longer manifestly covariant. We are currently addressing this problem.
\section*{Acknowledgements}
Roger Macleod had some insightful comments. T.H. would like to thank Dr. Arshad Momen for helpful discussions.

\section*{Appendix}
We use the Dirac Matrices $\gamma^{\mu}$ where
\[ \gamma^0=\left( \begin{array}{cc}
1 & 0 \\
0 & -1  \\
\end{array} \right)\quad\gamma^i=\left( \begin{array}{cc}
0 & \sigma^i \\
-\sigma^i & 0  \\
\end{array} \right)\eqno(A.1) \]
where $\sigma^i$ is a Pauli spin matrix. These satisfy the condition
\[ \left\lbrace \gamma^{\mu},\gamma^{\nu}\right\rbrace=2\eta^{\mu\nu} \quad \left(\eta^{\mu\nu}=diag(+---)\right)\eqno(A.2) \]
Furthermore, we employ the matrices
\[ \gamma^{5}=i\gamma^{0}\gamma^{1}\gamma^{2}\gamma^{3},\quad \sigma^{\mu\nu}=-\frac{1}{4}\left[\gamma^{\mu},\gamma^{\nu}\right]=\frac{i}{2}\epsilon^{\mu\nu\lambda\sigma}\sigma_{\lambda\sigma}\gamma^5\eqno(A.3) \]
It also is convenient to employ
\[ S^i=\left( \begin{array}{cc}
-\sigma^i & 0 \\
0 & \sigma^i  \\
\end{array} \right),\quad\Sigma^i=\left( \begin{array}{cc}
\sigma^i & 0 \\
0 & \sigma^i \\
\end{array} \right),\quad\alpha^i=\left( \begin{array}{cc}
0 & \sigma^i \\
\sigma^i & 0 \\
\end{array} \right)\eqno(A.4) \]
We employ the left derivative for Grassmann variables $\theta_i$
\[ \frac{d}{d\theta_i}\left(\theta_j\theta_k\right)=\delta_{ij}\theta_k-\delta_{ik}\theta_j \eqno(A.5) \]
Our convention for the Poisson Brackets are
\[ \left\lbrace F_1,F_2\right\rbrace=\left(F_{1,q}F_{2,p}+F_{2,q}F_{1,p}\right)-\left(F_{1,\psi}F_{2,\pi}+F_{2,\psi}F_{1,\pi}\right)\eqno(A.6a) \]
\[ \left\lbrace B_1,B_2\right\rbrace=\left(B_{1,q}B_{2,p}-B_{2,q}B_{1,p}\right)+\left(B_{1,\psi}B_{2,\pi}-B_{2,\psi}B_{1,\pi}\right)\eqno(A.6b) \]
\[ \left\lbrace B,F\right\rbrace=-\left\lbrace F,B\right\rbrace=\left(B_{,q}F_{,p}-F_{,q}B_{,p}\right)+\left(F_{,\psi}B_{,\pi}+B_{,\psi}F_{,\pi}\right)\eqno(A.6c) \]
where $F_i(B_i)$ are Grassmann odd(even) functions and we have the canonical variables $(q_i,p_i)$ and $(\psi_i, \pi_i)$ which are Bosonic and Fermionic respectively.

If $L=L(q_i,\dot{q}_i,\psi_i,\dot{\psi}_i)$ then
\[ p_i=\frac{\partial L}{\partial \dot{q}_i}\quad \pi_i = \frac{\partial L}{\partial \dot{\psi}_i}\eqno(A.8) \]
and
\[ H(q_i,p_i,\psi_i,\pi_i)=\dot{q}_{i}p_i+\dot{\psi}_{i}\pi_i-L \eqno(A.8) \]

\end{document}